\documentclass[aps,twocolumn,prl,showpacs,color,psfig,epsf]{revtex4}
\usepackage{amsmath}
\usepackage{color}
\usepackage{amsfonts}
\usepackage{epsf}
\usepackage{graphicx}
\begin{document}

\title{A stochastic model of supercoiling-dependent transcription}

\author{C. A. Brackley$^1$, J. Johnson$^1$, A. Bentivoglio$^3$, S. Corless$^2$, N. Gilbert$^2$, G. Gonnella$^3$, D. Marenduzzo$^1$}
\affiliation{$^1$SUPA, School of Physics and Astronomy, University of 
Edinburgh, Peter Guthrie Tait Road, Edinburgh, EH9 3FD, UK\\
$^2$MRC Human Genetics Unit, Institute of Genetics and Molecular
Medicine, The University of Edinburgh, Edinburgh EH4 2XU, UK\\
$^3$Dipartimento di Fisica, Universit\`a  di Bari \& INFN, Sezione di Bari, 
70126 Bari, Italy}

\begin{abstract}
  We propose a stochastic model for gene transcription coupled to DNA supercoiling, where we incorporate the experimental observation that polymerases create supercoiling as they unwind the DNA helix, and that these enzymes bind more favourably to regions where the genome is unwound. Within this model, we show that when the transcriptionally induced flux of supercoiling increases, there is a sharp crossover from a regime where torsional stresses relax quickly and gene transcription is random, to one where gene expression is highly correlated and tightly regulated by supercoiling. In the latter regime, the model displays transcriptional bursts, waves of supercoiling, and up-regulation of divergent or bidirectional genes. It also predicts that topological enzymes which relax twist and writhe should provide a pathway to down-regulate transcription. \\
  \textit{This article has been accepted for publication in Physical Review Letters, May 2016.}
\pacs{87.14.Gk,87.10.Mn}
\end{abstract}

\maketitle

The dynamics of transcription is a topic of paramount importance in cell biology and biophysics. It underpins the expression and regulation of genes, which is crucial to the development and function of 
all living organisms~\cite{alberts}. To initiate transcription of a gene, cells rely on the binding of proteins such as polymerases and transcription factors, to the promoter -- a DNA region shortly upstream 
of the gene~\cite{alberts}. As there are a finite number of copies of such proteins present within a cell, this process is inherently 
stochastic~\cite{transcriptionmodels1,transcriptionmodels2,vanoudernaarden,martin}.

In this work, we introduce a stochastic model of gene expression, which is fundamentally different from previous studies as it couples transcription to the dynamics of DNA twist and supercoiling. 
Supercoiling is a topological property of DNA, arising from its chiral nature~\cite{DNAtopology,jimnick,statisticalmechanicssupercoiling}. For B-DNA in its relaxed state, the two 
strands of the molecule wind around each other once approximately every 10 base pairs (bp), forming a right-handed double helix~\cite{DNAtopology}. Twisting DNA away from this relaxed state, so as 
to over or under-wind the double helix, introduces positive or negative supercoiling respectively; if large enough, this torsional strain can lead to writhing, or to DNA melting. Supercoiling thus 
refers to the difference in the linking number of the two DNA strands, ${\rm Lk}$, with respect to that in the relaxed state, ${\rm Lk_0}$; the global ${\rm Lk}$ is a topological invariant if the 
DNA is a loop or its ends are constrained~\cite{DNAtopology}, whereas it can vary for an open polymer whose ends can rotate. 

There are several observations which strongly suggest that DNA supercoiling is intimately related to transcription, and that it can regulate gene expression.
First, the ``twin supercoiled domain''
model~\cite{twinsupercoileddomain,twinsim,supercoilingregulation,stasiak,peter} is based on the long-standing theoretical observation that if rotation of the RNA polymerase and its associated transcription 
machinery is hindered, as is likely in the crowded intracellular environment, then gene transcription leads to the creation of positive supercoiling ahead of the tracking polymerase, and 
negative supercoiling in its wake. For every 10~bp or so which are transcribed, the linking number changes by $\Delta{\rm Lk}\approx+1$ ahead 
of the polymerase and by $\Delta{\rm Lk}\approx-1$ behind it. Recent experiments have quantified supercoiling by measuring the DNA binding affinity of psoralen, a chemical which 
intercalates preferentially where the double helix is under-wound~\cite{nick,kouzine}. These studies have shown that human chromosomes are organised into a set of supercoiling domains, whose structure is 
dramatically altered by inhibiting transcription.

Our model is based on these observations, and incorporates the dynamics of supercoiling into a stochastic description of gene regulation. It exhibits a switch between two regimes: one where gene expression is 
random, and one where it is tightly regulated by supercoiling. Within our framework, this switch is triggered, e.g., by increasing the amount of supercoiling injected during each transcription event. The dynamics 
in the supercoiling-regulated regime help explain a number of experimental observations, such as the existence of transcriptional bursts and the abundance of bidirectional genes in the genomes of many organisms.

We model the DNA as a 1D lattice with spacing $\Delta x\equiv l\sim 15$ bp, the size of an RNA polymerase~\cite{alberts,peter}. The DNA contains $n$ genes, each of size $\lambda$, whose promoters are located 
at positions $y_j$  ($j=1,\ldots,n$) on the DNA. Gene transcription is modelled as a stochastic process~\cite{SI}: at each time-step, for each of $N$ polymerases a gene is selected at 
random and is activated by the polymerase binding at the promoter with rate $k_{\rm on}$. Once a gene is activated, the polymerase travels along the gene body at a velocity $v$, so the position along the DNA 
of the $i-$th polymerase which is transcribing, say, the $j-$th gene is $x_i=y_j+vt_i$, where $t_i$ is the time since the polymerase was activated. The total time to transcribe any gene is then $\tau=\lambda/v$, 
after which the polymerase unbinds from the DNA and is free to transcribe another gene. [A simpler model where a static polymerase generates supercoiling without travelling is discussed in~\cite{SI}.]

We couple transcription to the local supercoiling density, $\sigma(x,t)=\left({\rm Lk}-{\rm Lk_0}\right)/{\rm Lk_0}$, where ${\rm Lk}$ is the local linking number at position $x$. We propose the following 
diffusive dynamics for $\sigma(x,t)$:
\begin{eqnarray}\label{model}
\frac{\partial \sigma(x,t)}{\partial t} & = & 
\frac{\partial}{\partial x}
\left[D\frac{\partial \sigma(x,t)}{\partial x} -J_{\rm tr}(x,t)\right],\\ 
\nonumber
J_{\rm tr}(x,t) & = & \sum_{i=1}^N J_i(t_i)\delta\mathbf{(}x-x_i(t_i)\mathbf{)}\xi_i(t), 
\end{eqnarray}
where $D$ is the effective diffusivity of supercoiling along DNA, and $J_{\rm tr}(x,t)$ is the local flux of supercoiling (Fig. 1) arising due to the transcription of any of the genes~\cite{SI}. We use 
periodic boundary conditions so that the overall level of supercoiling is conserved (this corresponds to modelling a DNA loop). In Eq.~(\ref{model}), $\xi_i(t)$ is set equal to 0 when the $i$-th polymerase is 
inactive, and to 1 when it is transcribing any of the $n$ genes. The modulus of the flux is $J_i=J_0\left(1+vt_i/l\right)$: it increases during transcription to model the fact that the positive supercoiling is 
racked up in front of the travelling polymerase. The sign of $J_i$ depends on the direction of gene transcription. Due to the observation that negative supercoiling can facilitate binding of RNA polymerases and 
transcription factors~\cite{bindtosupercoil,hatfield}, we further assume that $k_{\rm on}$ depends on the local value of $\sigma$ at the promoter, ${\sigma}_{\rm p}$. For simplicity we choose a linear coupling, 
$k_{\rm on}=k_0{\rm max}\left\{1-\alpha{\sigma}_{\rm p},0 \right\}$, where $k_0$ is the polymerase binding rate for $J_0=0$, and $\alpha$ quantifies the sensitivity to ${\sigma}_{\rm p}$. The linear dependence 
of $k_{\rm on}$ on ${\sigma}_{\rm p}$ is enough to give rise to highly non-linear dynamics. This is because the supercoiling created when a gene is switched on favours its own transcription, as well as that of 
upstream genes, whereas it hinders expression of the genes downstream. These chains of positive and negative feedbacks are at the basis of the non-linear transcription dynamics described below.

There are three main dimensionless parameters in the model. The first is the product of the transcription rate and the transcription time, $\Phi=(k_{\rm on}N/n)\tau$, which measures how often the gene is on. 
The second measures how fast supercoiling diffuses away between transcription events, $\Theta=(k_{\rm on}N/n)\lambda^2/D$. The third one is $\bar{J}/D$, and identifies the supercoiling generated near 
the promoter while the gene is active [$\bar{J}=J_0(1+\lambda/(2l))$ is the average supercoiling flux during transcription]. In~\cite{SI} we show that the average supercoiling at the promoter can be estimated in 
terms of these parameters as ${\sigma}_{\rm p}\sim-[(\Phi/(\Phi+1)] \bar{J}/(2D)$. [This estimate should be seen as a change from the baseline value of supercoiling, $\sim -0.05$ in bacteria.] Dimensional analysis 
further suggests that $\bar{J}\sim v\lambda$. The main question is then whether the average level of supercoiling generated triggers the positive feedbacks highlighted above; experiments suggest 
${\sigma}_{\rm p}\sim -0.01$ is enough to affect polymerase binding~\cite{bindtosupercoil,bindtosupercoil2}. 
What is the situation inside cells? The diffusion constant of supercoils within naked DNA is $D\sim0.1$~kbp$^2$/s or less~\cite{dynamicssupercoil}. Within bacteria, transcription rates are $\sim10$ RNA molecules 
per minute or above~\cite{Liang99}; considering a typical gene size of $1$~kbp and an elongation rate of 100 bp/s, we get ${\sigma}_{\rm p}\sim -0.3$. This suggests that supercoiling can be relevant for 
transcription in prokaryotes. In eukaryotes, transcription initiation is slower due to the need for several transcription factors to co-localise at a promoter; for example rates in yeast and humans are about 
10 and 1 transcripts per hour respectively~\cite{yeasttranscription,humantranscription}. Given that for eukaryotes $v\sim 25$bp/s, while $\lambda$ lies between 1.6 kbp (yeast) and 10 kbp (humans), we obtain 
${\sigma}_{\rm p}\sim -0.03$ (yeast) and ${\sigma}_{\rm p} \sim -0.13$ (humans). Because $D$ has not been measured for chromatin, these order-of-magnitude estimates should be viewed with caution, yet they
suggest supercoiling may affect polymerase initiation in eukaryotes as well~\cite{notetheta}.

\begin{figure}
  {\center \includegraphics[width=0.75\columnwidth]{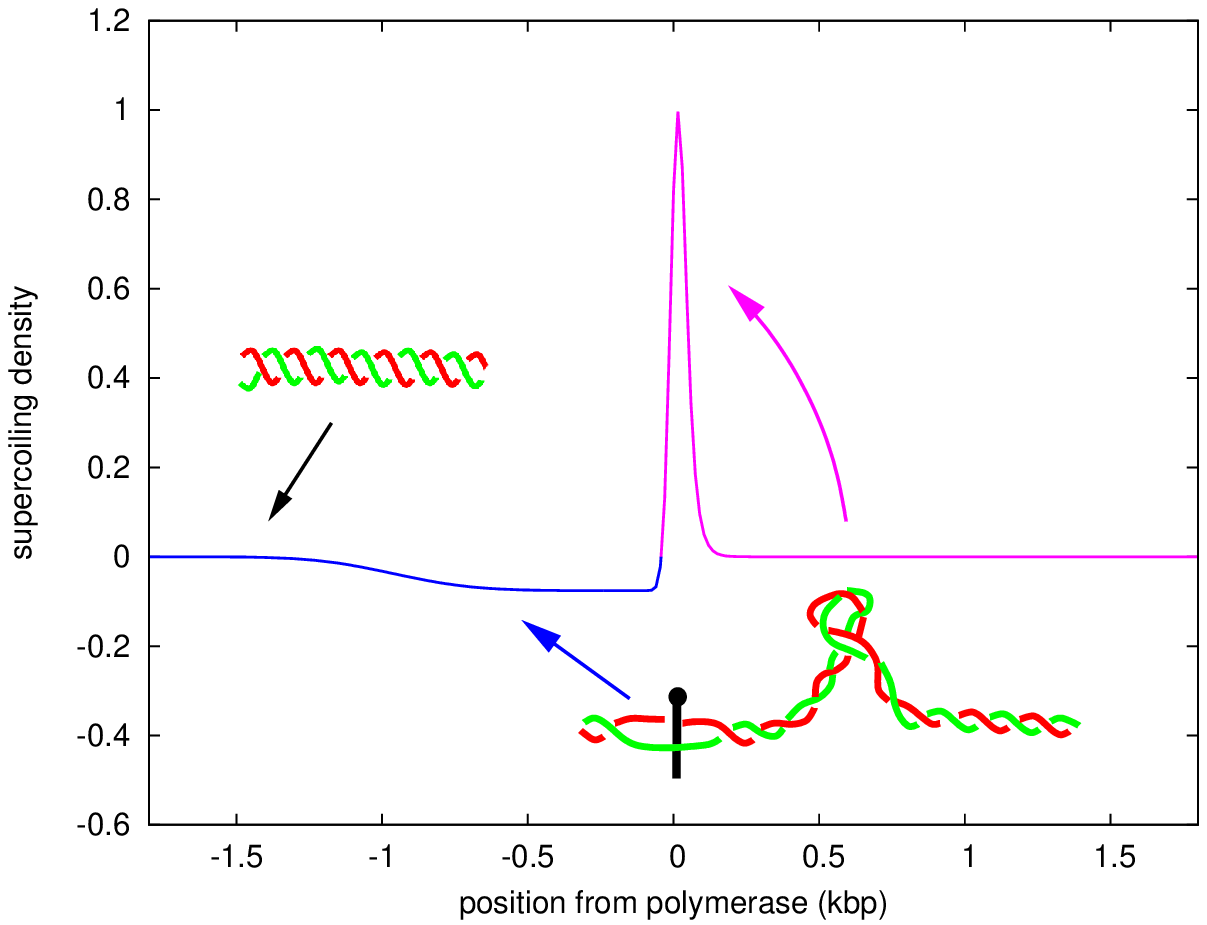} }
  
\caption{Schematic of the supercoiling density close to a transcribed gene, in the frame of reference of the travelling polymerase (see also Suppl. Movie 1). The RNA polymerase creates positive 
supercoiling (here speculatively depicted as right-handed writhe) ahead of the gene, while it generates negative supercoiling (here speculatively depicted as DNA unwinding) behind. The supercoiling profile is
obtained by solving Eq.~(\ref{model}) with $\bar{J}/D=1.7$, and other parameters as in~\cite{parameters}, except for $n=N=1$. The gene is switched on at time $t=0$, and the plot is for $t=\tau$; the transcription 
flux is here a regularised delta function~\cite{SI}.}
\label{fig1}
\end{figure}

Here and in what follows, we will choose parameters which are relevant to bacterial DNA~\cite{parameters}, and study how the system behaves upon varying $\bar{J}$. As discussed in~\cite{SI}, the results we report here are representative of the system's behaviour in general. We start by considering a 
case in which all genes are read in the forward direction, from left to right.  The genes are positioned randomly, but with the constraint that the distance between neighbouring genes is $>$1~kbp. For small 
$\bar{J}/D$, the typical values of $\sigma$ generated by transcription are modest (Fig. 2A, red curve, and Suppl. Movie 2): we call this the {\it relaxed} regime. The sequence of transcription events in this 
regime is well described by a Poisson process: any gene is read on average the same number of times, and the total number of transcription events is $\sim k_0 N T$ where $T$ is the total simulation time. 
As $\bar{J}/D$ increases, the flux of supercoiling injected by a polymerase becomes large enough to change the transcriptional dynamics significantly. Now the scale of variation of $\sigma$ is much larger 
(Fig. 2A, green curve, and Suppl. Movie 3): we call this the  {\it supercoiling-regulated} regime.
The value of ${\sigma}_{\rm p}$ is now large enough to affect $k_{\rm on}$ significantly, and this triggers bursts in transcription of the same gene, and waves of transcription (Fig. 2B, see also Suppl. Movies 3 
and 4). Genes are also no longer equally expressed: those with a large gap between them and the nearest upstream neighbour are up-regulated because they are less affected by the build-up of positive supercoiling 
during transcription (Fig. 2C).

As expected from the discussion above, the switch between the relaxed and supercoiling-regulated regimes is associated with a rise in overall transcription rate (Fig. 2D). It is also linked to a change in the 
nature of the time series describing the sequence of transcribed genes which becomes non-Poissonian and displays temporal correlations (due to bursting and waves of transcription). A useful way to quantify such 
a change is via the ``conditional entropy'' and ``mutual information''~\cite{conditionalentropy1,conditionalentropy2,informationtheorypaper} (definitions are given in~\cite{SI}). 
The conditional entropy is maximal and equal to $\log{(n)}$, if the transcription dynamics is a Poisson process (as is the case for $\bar{J}\to 0$), whereas it equals 0 in the 
limit of a maximally correlated process (e.g., when a single gene is repeatedly transcribed). Fig. 2D shows that the conditional entropy decreases with $\bar{J}/D$ in a sigmoidal way. The mutual information is 
a measure of the deviation of the observed joint probability distribution for successive transcription events, from that of a random process: for the case of Fig. 2, it is close to 0 for $\bar{J}=0$, and is higher 
in the supercoiling-regulated regime (Fig. S1~\cite{SI}). A semi-analytic theory of transcription bursts in a single gene model reproduces well the overall transcription rate of Fig. 2. A simplified mean field 
theory also shows that the switch is a crossover rather than a non-equilibrium phase transition, leads to the estimate for $\sigma_{\rm p}$ discussed above, and further suggests that supercoiling can affect 
transcription if $\bar{J}k_0\tau \alpha/(2D)\sim 1$ or larger~\cite{SI}.

\begin{figure}
\centerline{\includegraphics[width=.99\columnwidth]{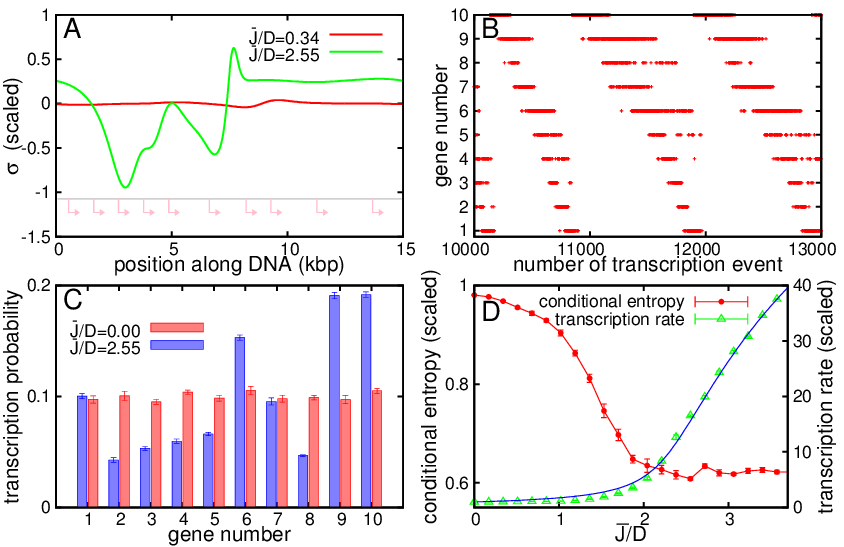}}
\caption{(A) Snapshots of $\sigma(x,t)$ in the relaxed regime (red, $\bar{J}/D=0.34$) and in the supercoiling-regulated regime (green, $\bar{J}/D=2.55$) for a 15~kbp DNA. (B) Portion of the time series of 
the sequence of transcribed genes for $\bar{J}/{D}=2.55$. A transcription wave can be seen as genes are transcribed preferentially in the order 10, 9, \ldots, 1, 10, \ldots (see Suppl. Movie 4; genes numbered 
from left to right). (C) Histograms showing gene transcription probabilities for $\bar{J}/D=0$ and $\bar{J}/D=2.55$ 
(average over 7 runs). The most transcribed genes for $\bar{J}/D=2.55$ are (in order) 10, 9, and 6. (D) Plot of the conditional entropy and the overall transcription rate (scaled by $k_0 N$; the blue line 
is the transcription rate predicted by the semi-analytic theory in~\cite{SI}). Gene positions for (A)-(D) are indicated in (A). Results for (C,D) were averaged over 7 runs.}
\label{fig2}
\end{figure}

In reality, genes can be encoded either in the forward or reverse strand of the DNA double helix~\cite{divergenttranscription1,divergenttranscription2,divergenttranscription3}, hence the supercoiling 
flux can be directed either way along the genome. To see how this affects our model, we study the case in which some of the genes are transcribed left to right, and others right to left (see Fig. 3). 
Figs. 3A and B show that in the supercoiling-regulated regime (large $\bar{J}/D$), some gene pairs are up-regulated together (see Fig. 3B). These are the divergent pairs (adjacent genes which point away 
from each other): when either is switched on, negative supercoiling is generated between the genes, which triggers further transcription in both. Within a given run, we normally observe transcription of 
a single divergent gene pair, where the selection mechanism is fluctuation-dependent (Fig. S4~\cite{SI}); within several runs, there is a ranking list of divergent pairs which depends quite subtly on gene 
position (Fig. 3B).  Transcription of convergent genes instead leads to a build-up of positive supercoiling, so is always strongly down-regulated.

In comparison to the case of genes which are all in the same direction, random orientations lead to a more marked peak in the mutual information, and to a sharper drop in the conditional entropy (Fig. 3C).
Divergent transcription also yields a larger overall transcription rate (again with respect to the case of parallel genes, see Fig. 3D). It is tempting to propose that this mechanism that markedly favours 
the transcription and co-expression of divergent pairs is amongst the reasons for the high abundance of such promoter pairs in the genomes of several organisms, including 
humans~\cite{divergenttranscription1,divergenttranscription2}. Furthermore, consistent with our model, divergent gene neighbours in yeast are often co-expressed, have low transcriptional noise and, 
importantly, are often associated with essential genes which tend to be highly expressed~\cite{yeastbioinformatics,notedrosophila}. 

\begin{figure}
\centerline{\includegraphics[width=0.99\columnwidth]{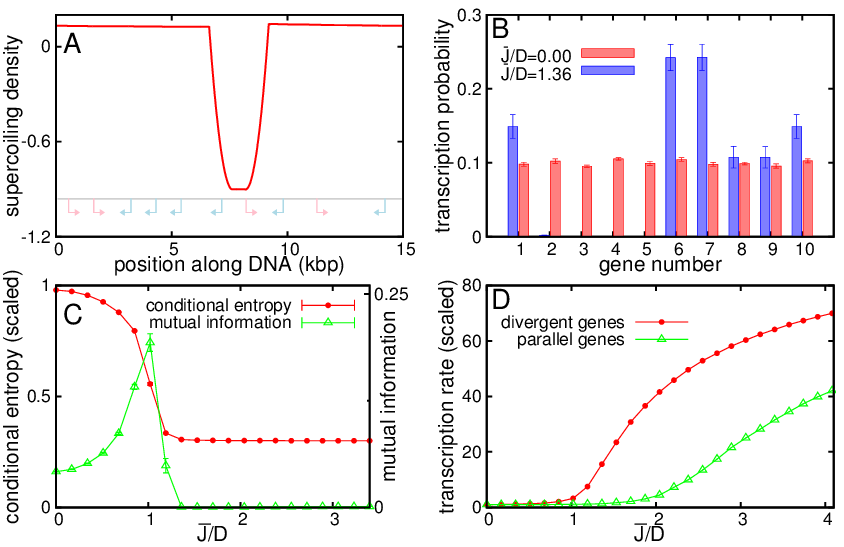}}
\caption{(A) Plot of the average value of $\sigma$ in the supercoiling-regulated phase for a 15~kbp DNA with forward and backward genes. (B) Histograms of 
transcription probabilities for the same system with $\bar{J}/D=0$ (red bars), and $\bar{J}/D=1.36$ (blue bars). The divergent pair 6, 7 is up-regulated because of the trail of parallel genes in front of 6. 
(C) Conditional entropy [scaled by $\log(n)$] and mutual information (averaged over 200 runs for the same gene positions as in A-B). (D) Overall transcription rate from all genes (scaled by $k_0 N$, 
averaged over 7 runs), for the single orientation arrangement of genes in Fig. 2, and for a divergent arrangement where the genes occupy the same region of DNA, but the first 5 are transcribed right to left.}
\label{fig3}
\end{figure}

Within a cell, the level of supercoiling is not conserved globally due to the presence of topological enzymes such as type I and type II topoisomerase, which can relax local supercoiling at a rate of 
$\sim0.1$-$1$~supercoil/s~\cite{toporate}. It is therefore of interest to include these enzymes in our model; the simplest way is through a non-conserved reaction term in Eq.~(\ref{model}), 
as follows,
\begin{equation}\label{topo}
\frac{\partial \sigma}{\partial t}  =  
\frac{\partial}{\partial x}
\left[D\frac{\partial \sigma}{\partial x} -J_{\rm tr}(x,t)\right]
-k_{\rm topo}\sigma,
\end{equation} 
where $k_{\rm topo}$ is a relaxation rate; this is associated with a length scale $l_{\rm topo}\sim \sqrt{D/k_{\rm topo}}$, over which supercoiling-mediated regulatory interactions are screened. Fig. 4 shows the 
effect of such enzymes in the set-up corresponding to Fig. 3. Divergent gene pairs are strongly up-regulated  if $k_{\rm topo}=0$, but for $k_{\rm topo}>0$ there is a dramatic down-regulation of transcription 
(Fig. 4A and 4B). This is accompanied by a rise in the conditional entropy (Fig. 4B); topoisomerases therefore rapidly lead to a loss of correlations in the transcription process.

\begin{figure}
\centerline{\includegraphics[width=0.99\columnwidth]{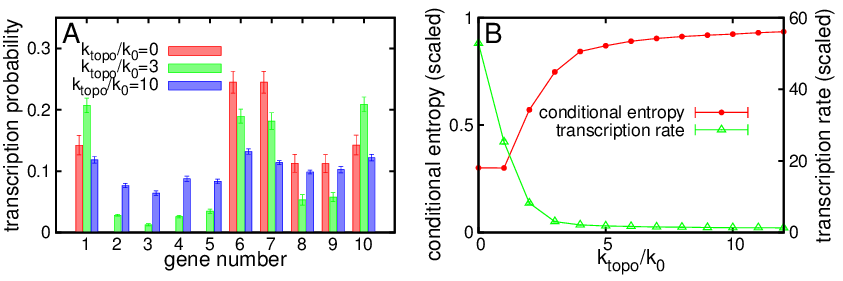}}
\caption{(A) Histograms of transcription probabilities of the 10 bidirectional genes in Fig. 3 (with $\bar{J}/D=2.55$), with different values of $k_{\rm topo}/k_0$. (B) Conditional entropy and transcription rate 
for the same system as A, as a function of $k_{\rm topo}/k_0$. Results were averaged over 6 runs.}
\label{fig4}
\end{figure}

In conclusion, we presented a dynamical model for supercoiling-dependent transcription, where a continuum description for the evolution of supercoiling is coupled to a stochastic transcriptional 
dynamics. Our model shows a crossover between two distinct regimes. When the supercoiling flux created as a polymerase transcribes a gene is small, transcription is a random process. When this flux 
is large, the dynamics become highly correlated. These correlations can be measured using the conditional entropy and mutual information of the transcriptional time series. For parallel genes, 
supercoiling drives transcriptional waves and bursts reminiscent of those observed in high-resolution dynamical experiments in both pro- and eukaryotes~\cite{burst1,burst2,burst3}. It also regulates 
gene expression, promoting the transcription of genes which have a larger gap separating them from their upstream neighbours. When considering genes with random orientations, transcription localises 
at divergent gene pairs, which are highly up-regulated. This is consistent with the observation that in yeast divergent gene pairs are often highly expressed essential genes~\cite{yeastbioinformatics}, 
and may explain the statistically surprising abundance of bidirectional promoters within mammals~\cite{divergenttranscription1,divergenttranscription2}. Finally, our theory predicts that including the 
action of topoisomerases, which locally relax supercoiling, down-regulates transcription: this agrees with the observation that inhibiting topo I can boost eukaryotic transcription rates 
{\it in vivo}~\cite{divergenttranscription1,supercoilingregulation}. Note that we disregard other important topological enzymes, such as the bacterial gyrase, whose role is to introduce, rather than to relax, 
negative supercoiling: such enzymes are known to promote transcriptional bursting~\cite{burst3}.

We foresee at least three major ways in which this work can be further pursued. First, we hope that our study will stimulate quantitative experiments measuring gene expression {\it in vitro}, where gene 
positions and directions can be controlled, e.g., via DNA editing.  Second, the model could be refined by comparison with high resolution psoralen data on supercoiling domains in both pro- and eukaryotes. 
Finally, it would be of interest to couple the dynamics of supercoiling to that of nucleosomes, which can at the same time create a barrier for supercoil diffusion, and localise twist and writhe.

CAB and DM acknowledge ERC for funding (ERC Consolidator Grant 648050, THREEDCELLPHYSICS).

\clearpage

\setcounter{figure}{0}
\renewcommand{\thefigure}{S\arabic{figure}}
\renewcommand{\theequation}{S\arabic{equation}}
\onecolumngrid
\section*{\Large A stochastic model of supercoiling-dependent transcription}

\section*{\Large Supplementary Information}
\begin{center}
C. A. Brackley$^1$, J. Johnson$^1$, A. Bentivoglio$^3$, S. Corless$^2$, N. Gilbert$^2$, G. Gonnella$^3$, D. Marenduzzo$^1$\\
\textit{$^1$SUPA, School of Physics and Astronomy, University of 
Edinburgh, Mayfield Road, Edinburgh, EH9 3JZ, UK\\
$^2$MRC Human Genetics Unit, Institute of Genetics and Molecular
Medicine, The University of Edinburgh, Edinburgh EH4 2XU, UK\\
$^3$ Dipartimento di Fisica, Universit\`a  di Bari \& INFN, Sezione di Bari, 
70126 Bari, Italy}
\end{center}

\vspace{1cm}

Here we provide additional details and results for the travelling polymerase model presented in the main text. We also discuss some variants of the model, including one in which polymerases are static. We also present some analytical results, mainly obtained within the simpler static polymerase model.

\section{Travelling polymerase model: additional details and variants considered}

The diffusive equation of motion for the supercoiling density, Eq.~(1) in the main text, can be motivated as follows. First, a good approximation for the free energy density of a supercoiled DNA with degree of supercoiling $\sigma$ is
\begin{equation}\label{freeenergy}
f=\frac{A}{2}\sigma^2
\end{equation}
where $A$ is a positive constant with appropriate dimensions~\cite{DNAtopology,statisticalmechanicssupercoiling}. In a DNA loop, or in a DNA region where the ends are fixed, or not free to rotate, the overall degree of supercoiling is fixed, therefore the appropriate dynamics for $\sigma(x,t)$ is that of model B~\cite{chaikinlubensky},
\begin{equation}\label{supercoileqdiff}
\frac{\partial \sigma(x,t)}{\partial t} = M \nabla^2 \frac{\partial f}{\partial \sigma} = MA \nabla^2 \sigma(x,t)\equiv D \nabla^2 \sigma(x,t),
\end{equation}
where $x$ is the 1D position along the DNA, $M$ is the mobility associated to the supercoiling density and $t$ is time. The term $\frac{\partial f}{\partial \sigma}$ is the analogue of the chemical potential in the ``standard'' model B dynamics (i.e., for binary mixtures~\cite{chaikinlubensky}.
As discussed in the text, Eq.~(\ref{supercoileqdiff}) means that the local degree of supercoiling diffuses, with effective diffusion coefficient $D=MA$. We note that Eq.~(\ref{freeenergy}) only holds for relatively small values of $\sigma$~\cite{DNAtopology,statisticalmechanicssupercoiling}. It would be possible, in principle, to improve this by choosing a free energy functional which better captures the free energy cost of supercoiling beyond the harmonic approximation. However, there are other aspects of the continuum model which break down for large $\sigma$; e.g., when the local supercoiling density becomes too negative, the polymerase would no longer indefinitely increase its affinity for the promoter, or when $\sigma$ is close to -1, so that the linking number of the DNA is close to 0, we would no longer expect transcription to create positive or negative supercoiling. Therefore, we feel it better to maintain the harmonic approximation which makes the model simpler, keeping in mind the caveat that it will break down for large values of supercoiling (say, $|\sigma|>1$).

It is useful here to make a few other technical remarks on the model. First, the supercoiling flux associated with transcription is proportional to a Dirac delta function (see Eq.~(1) of the main text). Since in reality a polymerase will have a finite footprint on the DNA, the Dirac delta function can also be substituted with its regularised representation, 
\begin{equation}\label{regularisation}
\delta(x) \to \exp{\left[-x^2/(4l^2)\right]/(2l\sqrt{\pi})},
\end{equation}
where $x$ is the argument of the Dirac delta function and $l$ is the regularised support of the flux. In the simulations for Figs. 2-4 in the main text (and also for Figs. S1-S6 below), the Dirac delta is substituted by a Kronecker delta $\delta_{x,0}$, hence regularisation occurs with $l\sim \Delta x$, the lattice spacing, which physically should be the size of a polymerase ($\sim$15 bp). In Fig. 1 in the main text we used the regularised delta of Eq.~(\ref{regularisation}), with $l=\Delta x$. 
Second, while the baseline model considers the case where the overall supercoiling integrates to 0, having an average non-zero supercoiling, $\sigma_0$ would not affect the results (provided that the dependence on $k_{\rm on}$ is on $\delta \sigma=\sigma-\sigma_0$).
Third, in the baseline model, a polymerase can engage on a gene as soon as its promoter is empty, i.e., when the previous polymerase has moved a single lattice spacing. In practice, a polymerase may need to be further away from the promoter before a second can initiate another transcription event. We have performed additional simulations, which suggests that while this fact quantitatively changes the overall transcription rate for a given value of the supercoiling flux, $\bar{J}$, the qualitative trends reported in the main text are preserved.
Fourth, in the travelling polymerase model reported in the main text the transcription flux  increases during transcription to model the fact that the positive supercoiling is racked up in front of the travelling polymerase. A constant $J_0$ would not capture the asymmetry between the higher positive supercoiling peak in front of the polymerase and the smaller negative supercoiling wake behind; we have seen that it would, however, lead to qualitatively similar physics.
Finally, when describing the model we imagine that the polymerase moves along the DNA. However, a similar level of torsional stress arises if the polymerase is immobilised and the DNA is reeled in to be transcribed~\cite{peter}.

In the main text, the switch between the relaxed regime and the supercoiling-regulated regime is described in terms of some information theory quantities~\cite{informationtheorybook,informationtheorypaper}, the conditional entropy and the mutual information, which we define mathematically here. These quantities are both defined in terms of a time series, here the sequence of the gene number transcribed over time, $\{i_q\}$, where e.g. $i_1$ is the index of the gene activated during the first transcription event, $i_2$ is that activated during the second etc. 

The ``conditional entropy'' of the transcription time series is defined as
\begin{equation}
S(\{i_q\})=-\sum_{i,j}p(i,j)\log{\left[p(i|j)\right]},
\end{equation}
where $p(i,j)$ is the joint probability of observing the transcription of gene $i$ followed by that of $j$ as consecutive events, while $p(i|j)$ is the conditional probability of gene $i$ being the next transcribed gene given that $j$ was the last to be transcribed. Note that in general $p(i,j)\ne p(j,i)$ (this is essentially due to the system being away from equilibrium). The conditional entropy $S(\{i_q\})$ is maximal, and equal to $\log{(n)}$, if the transcription dynamics is a Poisson process, as is the case when $\bar{J}=0$; it is instead equal to 0 in the limit of a maximally correlated process, for instance when a single gene is repeatedly transcribed.

The ``mutual information'' of the series of transcription events is defined as
\begin{equation}
I(\{i_q\})= \sum_{i,j}p(i,j)\log{\left[p(i,j)/(p(i)p(j))\right]},
\end{equation}
where $p(i)$ is the overall probability that gene $i$ is activated.
The mutual information is equal to 0 if $p(i,j)=p(i)p(j)$, i.e. for a succession of randomly chosen genes; its value therefore measures the divergence of the joint probability distribution for successive transcription events from that of a random process. In statistical mechanics systems it is often found that the mutual information peaks at or close to phase transitions~\cite{informationtheorypaper}, where correlations are maximal (however the definition of mutual information for thermodynamic systems is different~\cite{informationtheorypaper}).

\section{Travelling polymerase model: additional figures}

In this Section, we provide additional figures regarding the travelling polymerase model: these give more details and also demonstrate that the results shown in the main text are not qualitatively affected by different parameter choices.

Fig. S1 shows a plot of the mutual information for the case of genes with the same direction, as a function of $\bar{J}/D$ (same parameters and gene positions as in Fig. 2 of the main text): there is a shoulder accompanying the crossover between the uniform and the supercoiling-regulated regime.

Fig. S2 shows the transcription rate per gene as a function of $\bar{J}/D$ for the case of a single gene and a single polymerase (red curve). The curves are scaled by the transcription rate at $\bar{J}/D=0$. The single gene system can exhibit transcription bursts, but not waves, yet the overall transcription rate is similar to the overall transcription rate per gene in the case of genes oriented in the same direction (green curve, same data as in Fig. 2D).

For concreteness and simplicity, in the text we have chosen to vary only $J_0$ (hence $\bar{J}$), keeping other parameters constant (our parameter choice is relevant to bacterial DNA). This is sufficient because, as the supercoiling flux changes, this varies all three dimensionless parameters identified in the main text:$\bar{J}/{D}$, $\Theta$ and $\Phi$. However, it is also of interest to examine the quantitative effect of other parameters. Therefore in Fig. S3 we show the effect of varying $J_0$ (hence $\bar{J}$) when a different choice of the parameter $k_0$, the baseline polymerase binding rate, is used. We show both the transcription rate (Fig. S3A, scaled this time so that the value of 1 corresponds to all genes being constantly transcribed) and the mutual information (Fig. S3B) for the parallel array of genes in Fig. 2, where the value of the baseline polymerase binding rate without supercoiling, $k_0$, is varied by two orders of magnitude.

Fig. S3A shows that in all cases there is a crossover between a uniform regime and one where supercoiling upregulates transcription. Indeed, the curves show a similar crossover point when they are plotted as a function of $\bar{J}k_0\tau \alpha/D$ -- this scaling makes sense qualitatively since if genes are active less often (which is the case for smaller $k_0\tau$), then one expects that a larger flux is needed to enhance the transcription rate by a significant amount. Below we shall present a mean field theory which further motivates theoretically this scaling. The main effect of changing $k_0$ is that the crossover becomes sharper as $k_0$ decreases -- again, our mean field theory will provide an explanation for this observation.

Fig. S3B shows that the mutual information attains a similar value in the supercoiling-mediated regime, whereas the peak corresponding broadly to the crossover point is more visible for small $k_0$: again, this indicates that the crossover is sharper in that case. For the parameter range investigated, we further always find transcription waves in the supercoiling-regulated regime: these waves persist down to smaller values of $\bar{J}k_0\tau \alpha/D$ for small $k_0$.

Finally, Fig. S4 shows the transcription probability for different runs in the case of randomly oriented genes (positions as in Fig. 3 in the main text): the histograms show that in different runs, different gene pairs are upregulated. The system therefore shows multistability: once a gene pair is chosen, transcription is localised there for a long time (often for the whole run).

\begin{figure}[!h]
\includegraphics{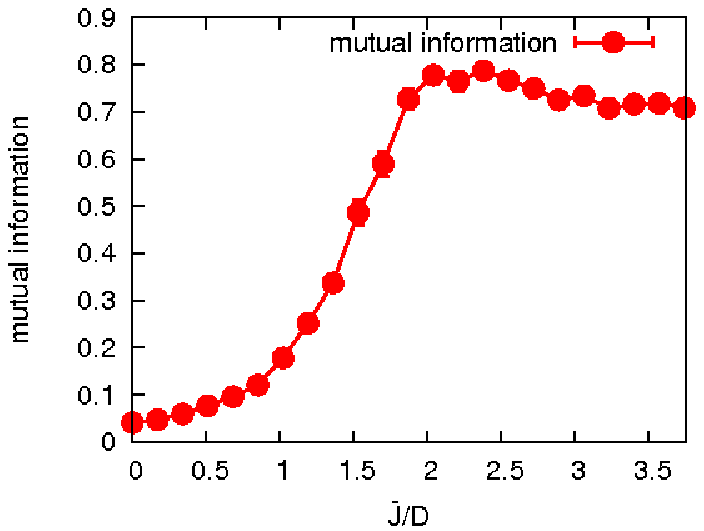}
\caption{Plot of the mutual information versus $\bar{J}/D$ for the transcription dynamics in Fig. 2 of the main text (averaged over 7 runs).}
\end{figure}

\begin{figure}[!h]
\includegraphics{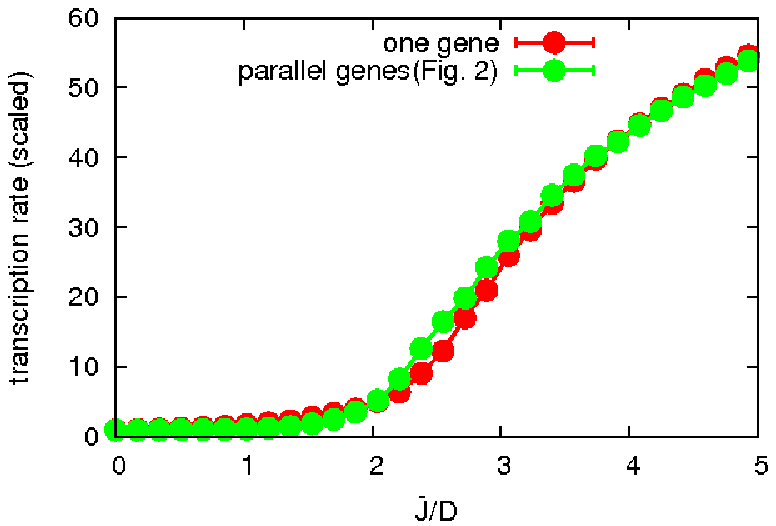}
\caption{Plot of the overall transcription rate (per gene) as a function of $\bar{J}/D$ for a model with only one gene ($N=n=1$; other parameters as in Fig. 2 of the main text) and for the case of genes oriented in the same direction (see Fig. 2 of the main text). The overall transcription rate is normalised with the expected value at $\bar{J}/D=0$ in both cases: the behaviour is very similar in the two cases.}
\end{figure}

\begin{figure}[!h]
  \begin{tabular}{cc}
    (A) & (B) \\
\includegraphics[width = .49\textwidth]{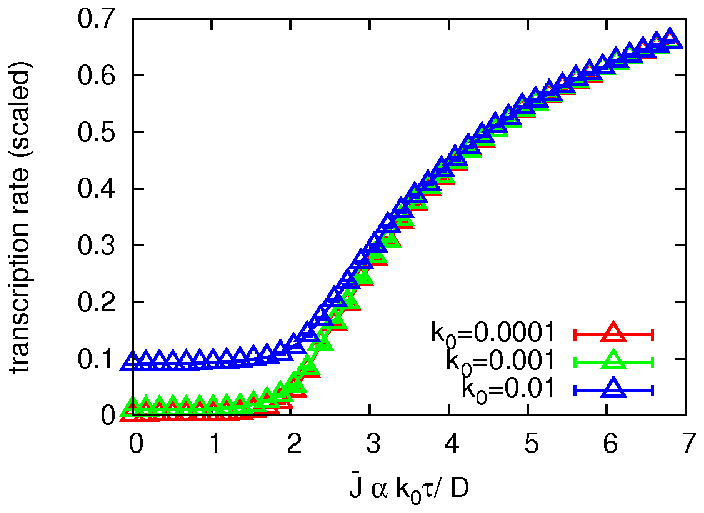} &
\includegraphics[width = .49\textwidth]{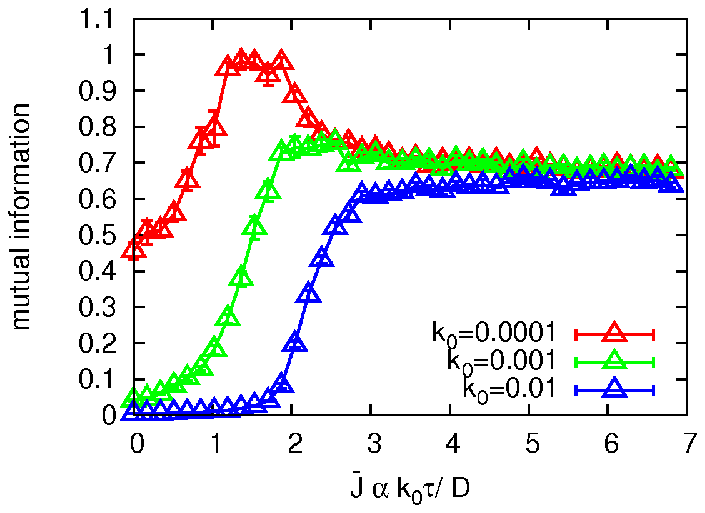}
  \end{tabular}
\caption{
(A) Plot of the overall transcription rate (per gene, and scaled such that it is 1 if all genes are transcribed constantly all of the time) as a function of $\bar{J}k_0\tau \alpha/D$ for the same gene arrangement as in Fig. 2 of the main text, and for different values of $k_0$, the polymerase binding rate in the absence of supercoiling (all other parameters as in Fig. 2; the value of $k_0$ is given in the legend, in s$^{-1}$, using the same mapping between simulation and physical units employed in the main text). The motivation for plotting the curves as a function of $\bar{J}k_0\tau \alpha/D$ comes from the mean field theory discussed in Section ``STATIC AND TRAVELLING POLYMERASE MODELS: MEAN FIELD THEORY, AND SCALING''). (B) Plot of the mutual information as a function of $\bar{J}k_0\tau \alpha/D$. The simulation with the smallest value of $k_0$ leads to a sharper increase of transcription rate with supercoiling flux (panel A), and to a better defined peak in the mutual information (panel B). All curves shown in (A) and (B) are averaged over 7 runs.}
\end{figure}

\begin{figure}[!h]
\includegraphics{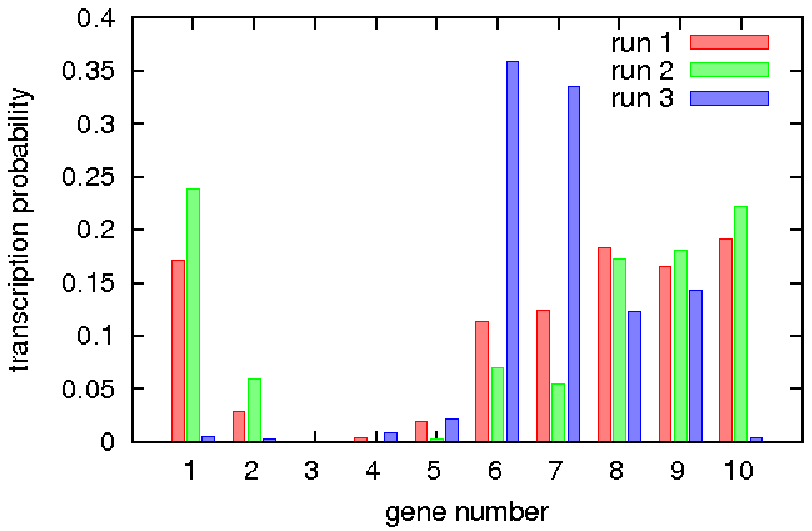}
\caption{Plot of the relative transcription probability for the 10 genes in Fig. 3, for $\bar{J}/D=1.02$ and for three different runs. Different divergent gene pairs are upregulated in different runs.}
\end{figure}

\clearpage 

\section{Static polymerase model: numerical results}

In this Section, we introduce and discuss a simpler model, where the polymerase does not travel along the gene body, but upon activating it introduces a supercoiling flux for a period of time $\tau$, over which the gene is transcribed. This model is less realistic biologically, but simpler to work with analytically (see next Section). We refer to this as the ``static polymerase model''.

The static polymerase model is analogous to the travelling polymerase model in that a stochastic dynamics of transcription is coupled to a diffusion-like equation for the supercoiling density, $\sigma$, i.e.,
\begin{equation}\label{modelSI}
\frac{\partial \sigma(x,t)}{\partial t} =  
\frac{\partial}{\partial x}
\left[D\frac{\partial \sigma(x,t)}{\partial x} -J_{\rm tr}(x,t)\right].
\end{equation}
In the static polymerase model, the form of the supercoiling flux during transcription is given by
\begin{equation}
J_{\rm tr}(x,t) = \sum_{i=1}^N J_i(t_i)\delta(x-x_i(t))\xi_i(t), 
\end{equation}
 where the sum is over the $N$ polymerases, and where $\xi_i(t)$ is set equal to 0 when the $i$-th polymerase is inactive, and equal to 1 when it is transcribing any of the $n$ genes. When $\xi_i=1$, the polymerase sits at the promoter of the gene it is transcribing: i.e., if the $i$-th polymerase is transcribing the $j$-th gene, then $x_i=y_j$. Finally, the modulus of the flux is $J_i=\pm J_0$, where the sign depends on gene direction as in the main model. Therefore, this represents the limit where $v=0$ in the travelling polymerase model. It is worth noting that in the static polymerase model, since the polymerase always sits at the promoter, there can be only one polymerase transcribing a gene at a given time. 

The dependence of $k_{\rm on}$ on supercoiling is taken to have a similar functional form as in the travelling polymerase model (see main text),
\begin{equation}\label{konstatic}
k_{\rm on}=k_0{\rm max}\left\{1-\alpha{\sigma}_{\rm p, x_0},0 \right\},
\end{equation}
however, in this case, ${\sigma}_{\rm p, x_0}$ is computed a distance $x_0$ upstream of the promoter of a given gene. Taking $x_0=0$ as in the travelling polymerase model would lead to artifical results, as, for instance, in a model with a single gene (at $x=0$), $\sigma(0,t)=0$ for symmetry when the gene is off (see next Section).

Fig. S5 and Fig. S6 show the behaviour of the static polymerase model (for $x_0=5 \Delta x$) as a function of $J_0/D$ for the case of genes oriented in the same direction (Fig. S5) and for divergent transcription (Fig. S6). The trends observed are qualitatively in agreement with those found with the travelling polymerase model (see Fig. 2 and Fig. 3 in the main text).

\begin{figure}[!ht]
  \begin{tabular}{cc}
    (A) & (B) \\
    \includegraphics[width = .49\textwidth]{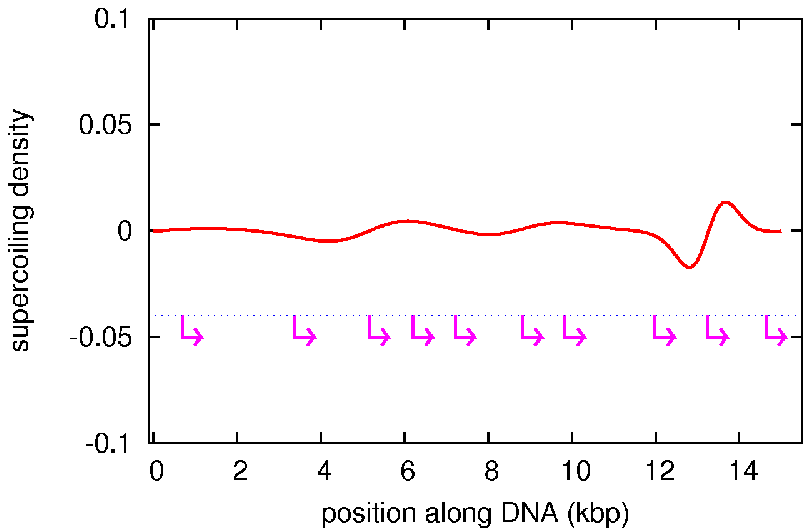} & \includegraphics[width = .49\textwidth]{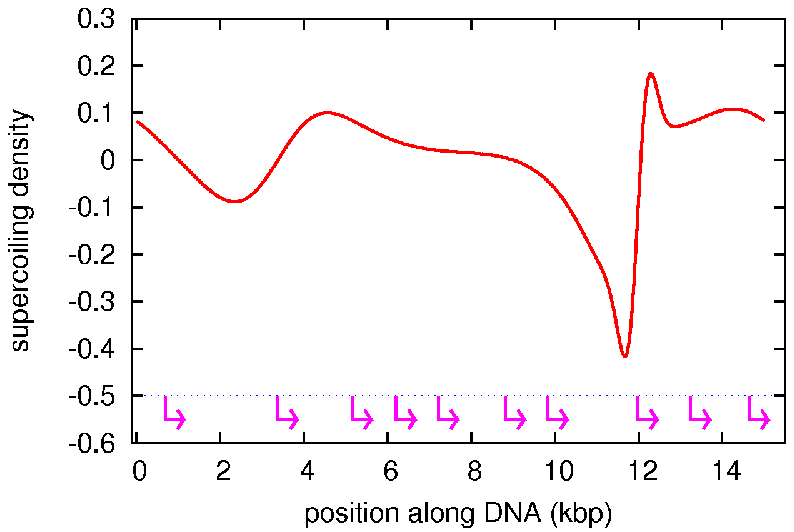} \\
    (C) & (D) \\
    \includegraphics[width = .49\textwidth]{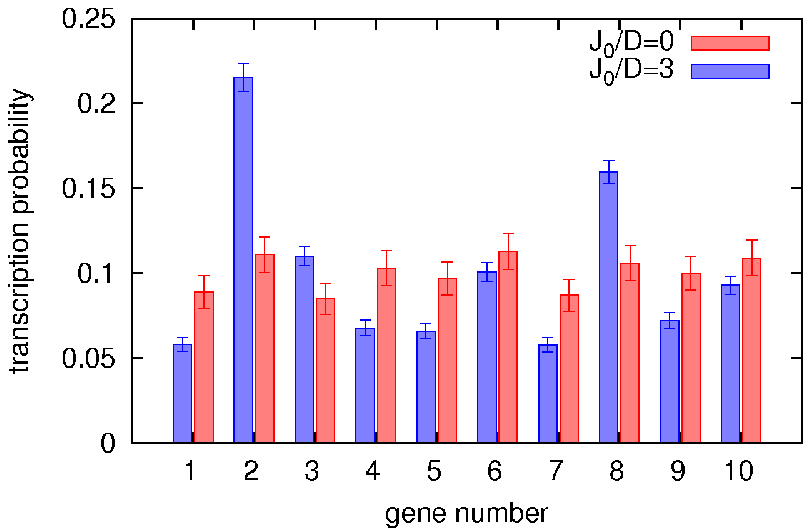} & \includegraphics[width = .49\textwidth]{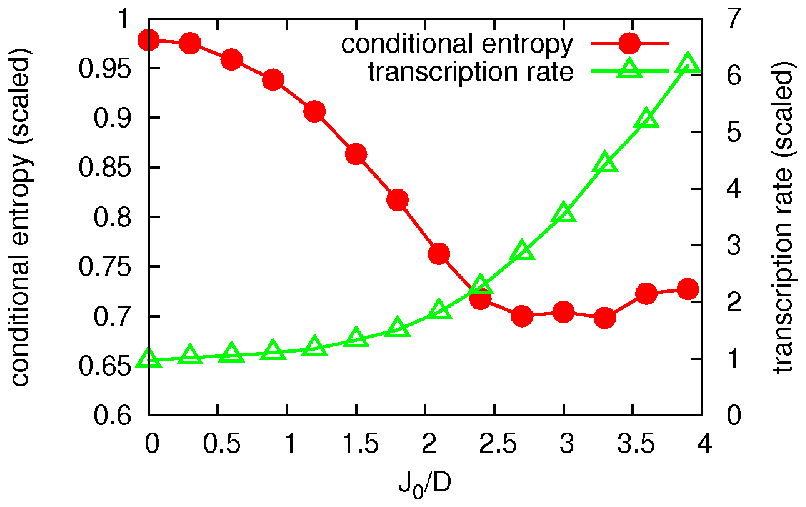}
    \end{tabular}
\caption{Static polymerase model with genes in the same direction. 
(A,B) Plots (snapshots) of the local supercoiling density in the relaxed
phase (A, $J_0/D=0.3$) and in the supercoiling-regulated regime 
(B, $J_0/D=3$) for a 15~kbp DNA. The gene positions are indicated in pink.
(C) Histograms showing the transcription probability of the genes in A and B
for $J_0=0$ (red) and $J_0/D=3$ (blue). Runs initialised with different seeds 
lead to the same set of genes being up or downregulated, for large $J_0/D$ --
in (C) the most transcribed genes are (in order) 2, 8, 3, 6. Upregulated
genes tend to be those which have a gap between them and the upstream neighbour,
so that they are not inhibited by the wave of positive supercoiling triggered
by its transcription. (D) Plot of the conditional entropy and the overall 
transcription rate (averaged over 17 runs for the same gene positions as 
in A-C).}
\end{figure}

\begin{figure}[!ht]
  \begin{tabular}{cc}
    (A) & (B) \\
    \includegraphics[width = .49\textwidth]{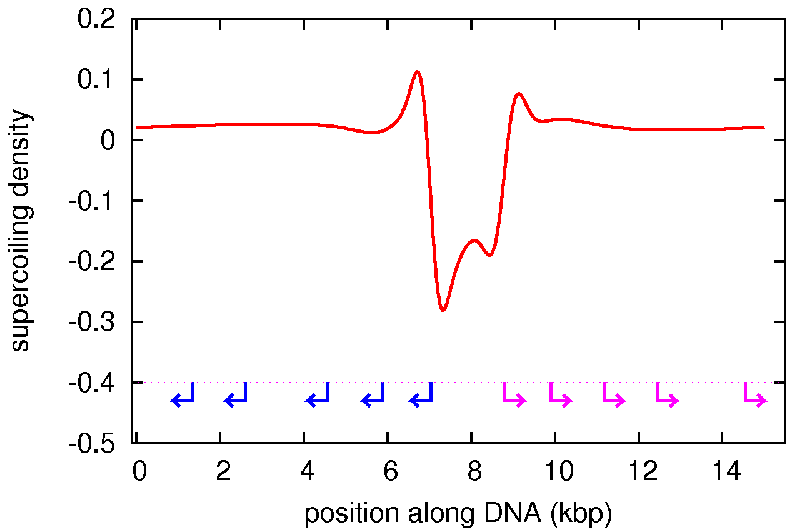} & \includegraphics[width = .49\textwidth]{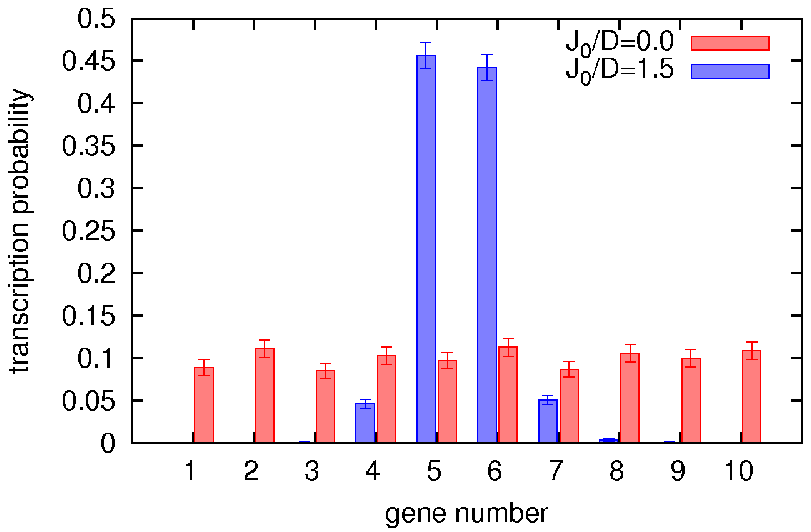} \\
    (C) & (D) \\
    \includegraphics[width = .49\textwidth]{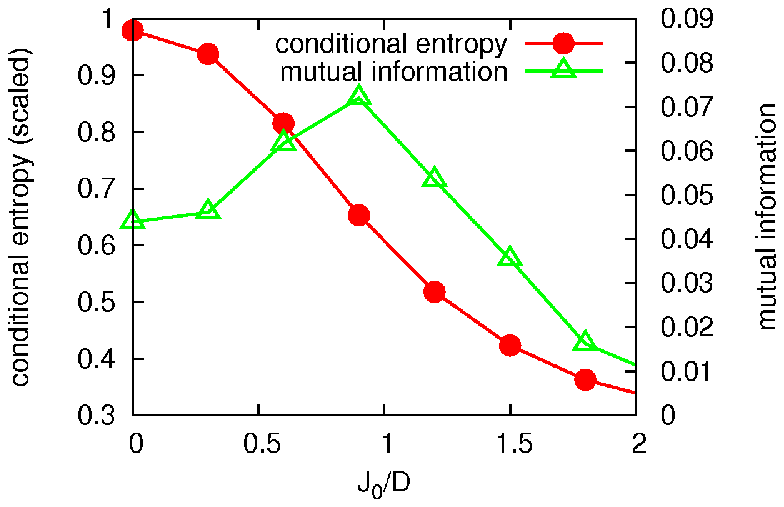} & \includegraphics[width = .49\textwidth]{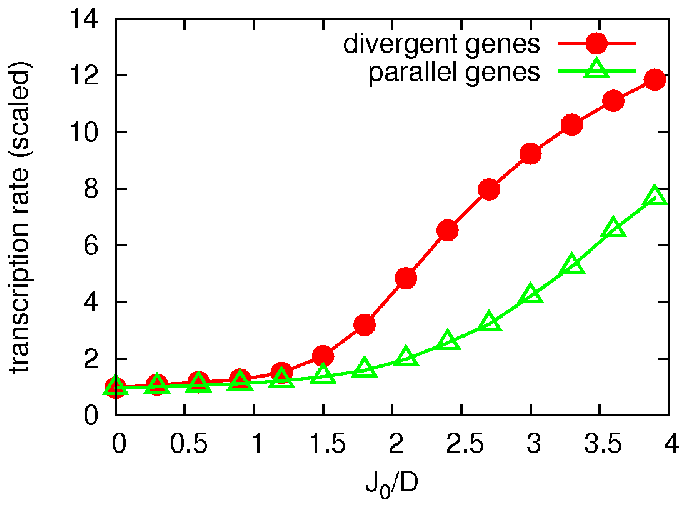}
    \end{tabular}
\caption{Static polymerase model with genes oriented along different directions. 
(A) Plot of the local supercoiling density (snapshot)
in the supercoiling-regulated phase for a 15~kbp DNA with forward and backward genes (positions indicated in pink and blue respectively). (B) Histograms of transcription probabilities for the same system with $J_0/D=0$ (red bars), and $J_0/D=1.5$ (blue bars).
(C) Conditional entropy [scaled by $\log(n)$] and mutual information (averaged over 30 runs) for the same gene positions as in A-B.
(D) Overall transcription rate (scaled by $k_0 N T$, and averaged over 30 runs) for the system in A, and for another system where genes are in the same positions but are all read in the forward direction.}
\label{figS3}
\end{figure}

\clearpage

\section{Static polymerase models: exact results}\label{analytics}

In this Section we obtain some exact results and scaling relations; we will work within the static polymerase model, but in the next Section we will also apply them to the travelling polymerase model.

We begin by considering the static polymerase model, where there is a single gene. We start from the equation for $\sigma(x,t)$, and imagine that the gene is on:
\begin{equation}\label{supercoilingeq1}
\frac{\partial \sigma(x,t)}{\partial t} = 
\frac{\partial}{\partial x}\left[D\frac{\partial\sigma(x,t)}{\partial x} - J_0\delta(x)\right],
\end{equation}
where we use the boundary condition that $\sigma(0,t)=0$, and consider no flux boundaries (so that the overall supercoiling is fixed; we solve the equations on an infinite domain, so this implies $\frac{\partial \sigma}{\partial x}=0$ at $x\rightarrow\pm\infty$). 
In steady state ($\frac{\partial \sigma(x,t)}{\partial t}=0$), the solution of Eq.~(\ref{supercoilingeq1}) is given by
\begin{equation}\label{solution1}
\sigma(x)=\frac{J_0}{2D}{\rm sgn}(x),
\end{equation}
where ${\rm sgn}(x)$ is the sign function, so $\sigma=\frac{J_0}{2D}$ for positive $x$, and $\sigma=-\frac{J_0}{2D}$ for negative $x$. This solution shows that the typical value of the supercoiling density is $|\sigma|\sim J_0/D$ (however it is only accurate for a gene which is always on). 

It is also of interest to examine how the solution evolves in time to yield Eq.~(\ref{solution1}) at steady state. To address this, we consider an initial condition with  $\sigma\equiv 0$, and we imagine that the gene is switched on at time $t=0$. Then, while the gene is switched on, the Laplace transform of $\sigma(x,t)$, which we shall call $\hat{\sigma}(x,s)$, with
\begin{equation}
\hat{\sigma}(x,s)\equiv \int_0^{\infty} dt \exp(-st)\sigma(x,t),
\end{equation}
satisfies the following equation
\begin{equation}\label{supercoilingeq2}
D\frac{\partial^2 \hat{\sigma}}{\partial x^2} - s\hat{\sigma} = -J_0 \frac{\delta'(x)}{s},
\end{equation}
where $\delta'(x)$ represents the derivative of the Dirac delta function. 

One way to solve Eq.~(\ref{supercoilingeq2}) is to observe that the Green's function, i.e. the solution of 
\begin{equation}
D\frac{\partial^2 g(x,x')}{\partial x^2} - sg(x,x') = \delta(x-x'),
\end{equation}
which decays to 0 at $|x-x'| \to \infty$, is given by
\begin{equation}\label{greenfunction}
g(x,x')=\frac{\exp\left(-\sqrt{\frac{s}{D}}|x-x'|\right)}{2\sqrt{Ds}}.
\end{equation}
Then, the solution of Eq.~(\ref{supercoilingeq2}) is
\begin{eqnarray}
\hat{\sigma}(x,s) & = & \int_{-\infty}^{+\infty} dx' \, g(x,x') \left[-J_0\frac{\delta'(x')}{s}\right] \\
\nonumber
& = & \frac{J_0}{2Ds}\exp\left(-\sqrt{\frac{s}{D}}|x|\right){\rm sgn}(x).
\end{eqnarray}
In real space, the solution is found by inverse Laplace transform; at time $t=\tau$, when transcription stops in our model, it is given by
\begin{equation}\label{supercoilingeq3}
\sigma(x,\tau)=\frac{J_0}{2D}{\rm erfc}\left(\frac{|x|}{2\sqrt{D\tau}}\right)
{\rm sgn}(x),
\end{equation}
where ${\rm erfc}$ is the complement of the error function. This solution tends to Eq.~(\ref{solution1}) when $\tau\to\infty$; it also shows that, while the gene is on, again the typical value of supercoiling density \textit{in the neighbourhood of the promoter} is $\sim J_0/D$. 

After the gene is switched off the supercoiling density satisfies the diffusion equation, 
\begin{equation}\label{supercoilingeq4}
\frac{\partial \sigma(x,t)}{\partial t} = D
\frac{\partial^ 2}{\partial x^2}\sigma(x,t),
\end{equation}
with the initial condition that $\sigma(x,\tau)$ is as given by Eq.~(\ref{supercoilingeq3}). The solution can be written as
\begin{equation}\label{supercoilingeq5}
\sigma(x,t)=
\int_{-\infty}^{\infty} dx' \, \frac{\exp\left[-\frac{(x-x')^2}{4Dt}\right]}{\sqrt{4\pi Dt}}\frac{J_0}{2D}{\rm erfc}\left(\frac{|x'|}{2\sqrt{D\tau}}\right){\rm sgn}(x'),
\end{equation}
where for simplicity we have shifted time so that the gene switches off at time $t=0$ and the solution holds for $t\ge 0$. Eq.~(\ref{supercoilingeq5}) can be used to infer that $\sigma(0,t)\equiv 0$ (for the static polymerase model), and $\sigma(x,t)\sim t^{-3/2}$ for large $t$ and for $x\ne 0$.

\section{Static and travelling polymerase models: mean field theory, and scaling}\label{analyticsmeanfield}

We now use the results obtained from the last Section to build a simple mean field theory for our model.

We start from the observation that, within the static polymerase model, the on rate for RNA polymerase, $k_{\rm on}$, depends on the extent of negative supercoiling upstream of the promoter (at $x_0<0$), according to the formula (see main text and Eq.~(\ref{konstatic})),
\begin{equation}\label{kon}
k_{\rm on}=k_0\left[1-\alpha\sigma(x_0,t)\right],
\end{equation}
where, since this is always positive, we do not need the ${\rm max}$ function as in the main text. 

We propose a simple mean field theory, where the value of $\sigma(x_0,t)$ is replaced with the average supercoiling profile over the whole simulation, $\bar{\sigma}(x_0)$. An equation for $\bar{\sigma}$ can be written down by finding the steady state solution of Eq.~(\ref{supercoilingeq1}) when the flux is replaced by its average $J_0\delta(x) \frac{k_{\rm on}\tau}{k_{\rm on}\tau+1}$, where $\frac{k_{\rm on}\tau}{k_{\rm on}\tau+1}$ is the fraction of time that the gene is on (this last formula can be obtained by realising that the polymerase has an on rate equal to $k_{\rm on}$ and an effective off rate equal to $1/\tau$). If we do this, we find that 
\begin{equation}\label{solution2}
\bar{\sigma}(x_0)=-\frac{k_{\rm on}\tau}{k_{\rm on}\tau+1}\frac{J_0}{2D}.
\end{equation}
We should note that this solution, as the previous ones, works for open, no flux, boundary conditions (our simulations instead have periodic boundary conditions, but the scaling of $\bar{\sigma}$ does not change).

We can now plug in this expression for $\bar{\sigma}$ in Eq.~(\ref{kon}), to get a self-consistent equation, similar in spirit to a mean field theory,
\begin{equation}\label{kon2}
k_{\rm on}=k_0\left[1-\alpha\bar{\sigma}(k_{\rm on})\right] \sim
k_0\left[1+\alpha \frac{k_{\rm on}\tau}{k_{\rm on}\tau + 1}\frac{J_0}{2D}\right].
\end{equation}
Eq.~(\ref{kon2}) has a solution which depends smoothly on $\frac{J_0}{D}$: in other words, there should be no discontinuity in the transcription rate (proportional to $k_{\rm on}$, see below) as a function of $J_0$. Another way to understand this is to realise that Eq.~(\ref{kon2}) is essentially equivalent to the mean field equation for the magnetisation versus temperature in the Ising model in the presence of a non-zero magnetic field (the $k_0$ term): it is well known that this equation in this case describes a smooth crossover and no thermodynamic phase transition. 

While we have derived our mean field equation, Eqs.(\ref{solution2}) and ~(\ref{kon2}) for the static polymerase model, numerically we found that Eq.~(\ref{solution2}) also applies well for the travelling polymerase model, with $J_0$ replaced by $\bar{J}$, the average supercoiling flux during transcription. Specifically, for the travelling polymerase model, the average supercoiling density at the promoter, which we call $\bar{\sigma}_{\rm p}$, is given by
\begin{equation}\label{solution3}
\bar{\sigma}_{\rm p} = -\frac{k_{\rm on}\tau}{k_{\rm on}\tau+1}\frac{\bar{J}}{2D} = -\frac{\Phi}{\Phi+1}\frac{\bar{J}}{2D},
\end{equation}
where $\Phi=k_{\rm on}\tau$ is one of the dimensionless numbers introduced in the main text, for $N=n=1$. Eq.~(\ref{solution3}) is used in the main text to estimate the supercoiling densities at promoters in bacteria, yeast and human cells.

By plugging Eq.~(\ref{solution3}) into Eq.~(\ref{kon}), we can find an explicit expression for $k_{\rm on}$ in our mean field theory, which is given by
\begin{eqnarray}\label{solution4}
k_{\rm on} \tau & = & \frac{h+\sqrt{h^2+4k_0\tau}}{2} \\ \nonumber
h & = & k_0\tau\left(1+\frac{\alpha \bar{J}}{2D}\right)-1.
\end{eqnarray}
The overall transcription rate $k_{\rm t}$ (of the single gene considered up to now in the simplified theory) can be estimated as follows,
\begin{equation}\label{transcriptionrate}
k_{\rm t} = \frac{k_{\rm on}}{1+k_{\rm on}\tau}
\end{equation}
where the correction $\frac{1}{1+k_{\rm on}\tau}$ takes into account the fact that the maximum transcription yield per gene is equal to $1/\tau$, when the polymerase is transcribing the gene at all times. Fig. S7 shows some examples of the overall transcription rate $k_{\rm t}$, for different values of $k_0\tau$. As anticipated when analysing the static polymerase model, for any $k_0\ne 0$, there is no discontinuity in the transcription rate, so that the switch between uniform and supercoiling-regulated regime is a crossover. The only limit in which this would become a true nonequilibrium transition is if $k_0\to 0$, while keeping the product $\bar{J}\alpha k_0\tau/D$ constant. Eqs.~(\ref{solution4}) and~(\ref{transcriptionrate}) also highlight a useful criterion to determine when supercoiling starts to significantly affect transcriptional rate (hence transcription): this occurs when 
\begin{equation}\label{threshold}
\frac{\bar{J}\alpha k_0\tau}{2D} \sim 1.
\end{equation}
In other words, the value of $\bar{J}/{D}$ (which is the parameter varied in the main text) at which we should expect the crossover between the uniform and the supercoiling-dominated regime is equal to $2/(\alpha k_0\tau)$. Eq.~(\ref{threshold}) also motivates the scaling used in Fig. S3.

Note that, as is the case in general for mean field approximations, the assumption that $k_{\rm on}$ depends on the {\it average} supercoiling profile, $\bar{\sigma}$, is only appropriate when the supercoiling profile does not vary too much in time, so that the instantaneous profile for $\sigma$ is close to the average one. This is the case when there is not enough time for the supercoiling to diffuse away in between transcription events. The physical dimensionless parameter determining when this is the case, in the travelling polymerase model, is $\Theta=\frac{k_{\rm on}\lambda^2}{D}$. If $\Theta$ is small, then diffusion is fast and while the gene is off the supercoiling is much smaller than the average value, and our mean field theory is not valid. 

Fortunately, even when $\Theta$ is relatively small (Fig. S8, where the minimum value of $\Theta$ is $\sim$0.44) our numerical results suggest that the value of $\sigma$ at the promoter, $\sigma_{\rm p}$, at the moment when the gene is switched on (which is the relevant value to use in Eq.~(\ref{kon})), depends on $k_{\rm on}$ linearly for small $k_{\rm on}$, so that the same qualitative considerations apply as in our simplified mean field theory (i.e., the system displays a crossover rather than a phase transition as $\bar{J}/D$ is increased). We can further perform a simulation to find the value of $\sigma_{\rm p}$ as a function of $k_{\rm on}$ (kept constant for each simulation, see Fig. S8 and its caption). We can then fit the resulting data with the following functional form,
\begin{equation}\label{fit}
|\sigma_{\rm p}| = \frac{ak_{\rm on}}{bk_{\rm on}+1}
\end{equation}
where $a$ and $b$ are positive constants determined via fitting (see Fig. S8). At this point, we can follow the procedure described above, where Eq.~(\ref{fit}) is plugged into Eq.~(\ref{kon}) to yield a semianalytical estimate for $k_{\rm on}$: this is an improvement with respect to the mean field estimate, Eq.~(\ref{solution4}). In a system with one polymerase and one gene, the rate $k_{\rm on}$ determined self-consistently via Eq.~(\ref{kon}) gives the overall transcription rate $k_{\rm t}$ by using Eq.~(\ref{transcriptionrate}). 
For a system with $N$ polymerases and $n$ genes, substituting $k_{\rm on}$ with $k_{\rm on}N/n$ we obtain the predicted transcription rate per gene. This rate is a good approximation of the transcription rate per gene in the case of genes oriented along the same direction (see the blue curve in Fig. 2D in the main text).

\begin{figure}[!h]
\includegraphics{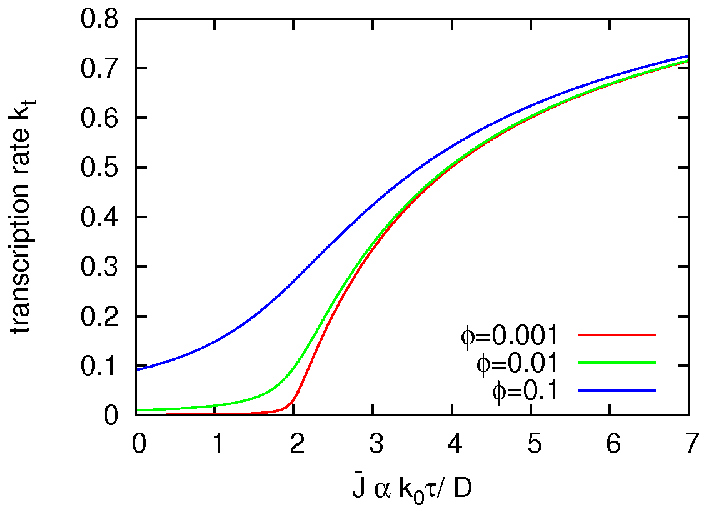}
\caption{Plot of the transcription rate, found by using Eq.~(\ref{solution4}) and Eq.~(\ref{transcriptionrate}), for $\alpha=100$ (as in the main text), and different values of $k_{\rm on}\tau$ (see legend).}
\end{figure}

\begin{figure}[!h]
\includegraphics{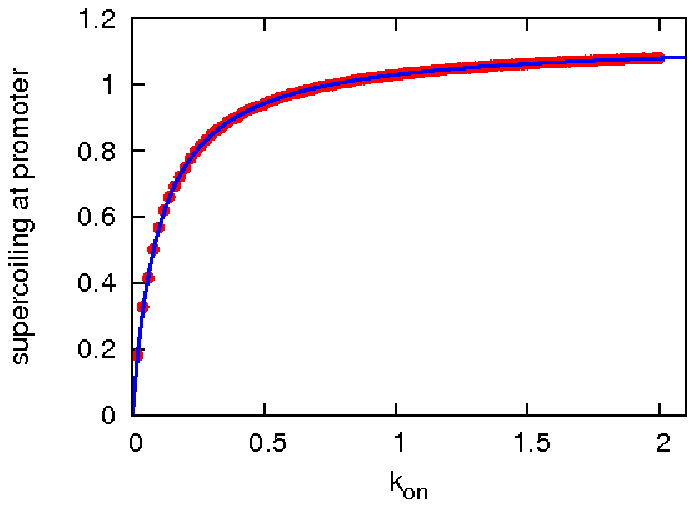}
\caption{(A) Plot of the local supercoiling density (absolute value) at the promoter as a function of $k_{\rm on}$ for a single gene, on a lattice of size 1000 $\Delta x$ (with periodic boundary condition). To make this plot we run our simulations with $\alpha=0$ so that $k_{\rm on}$ can be fixed as an input. The fit is to Eq.~(\ref{fit}), and the resulting parameters are $a\sim 11.18\pm 0.02$ and $b=9.85\pm 0.02$. This simulation was performed with $\bar{J}/{D}=2.55$; in order to get the transcription rate as a function of $\bar{J}$ we further assumed a linear dependence of $\sigma_{\rm p}$ on $\bar{J}$ overall (as in Eq.~\ref{solution2}).}
\end{figure}

\clearpage

\section*{Captions for Supplementary Movies}

Below are the captions for Suppl. Movies 1-4.\\

{\bf Suppl. Movie 1}: An example of dynamics corresponding to Fig. 1 of the main text, showing detail of one transcription event (note that the flux here is discretised as in Eq.~(S3), with $l=1$). \\

{\bf Suppl. Movie 2}: An example of dynamics corresponding to the relaxed regime, with $\bar{J}/D=0.34$, where all genes are oriented along the same directions. All other parameters are as in Fig. 2 in the main text. The bottom panel shows the number of the transcription event versus the number of transcribed gene. It can be seen that genes are transcribed in a random sequence. Here and in the following Suppl. Movies 3 and 4, the flux here is discretised as a Kronecker delta, as in Figs. 2-4 of the main text. Also note that, in order to cover the whole dynamics, the frame rate is too fast to resolve single transcription events.\\

{\bf Suppl. Movie 3}: An example of dynamics corresponding to the supercoiling-regulated regime, with $\bar{J}/D=1.7$. All other parameters as in Suppl. Movie 2, or Fig. 2 in the main text. The bottom panel shows the histograms of transcription events for each gene. It can be seen that some of the genes are transcribed more than others; the speeded up dynamics also show a transcription wave going from right to left. \\

{\bf Suppl. Movie 4}: As Suppl. Movie 3, but now the bottom panel shows the number of transcription events as a function of number of transcribed gene. The bottom plot highlights the transcription wave, as genes are more likely to be transcribed in the order 10, 9, \ldots, 1, 10, \ldots. \\



\begin{thebibliography}{99}
\bibitem{notetheta} These are average values, but typically $\sigma$ remains close to this provided  that $\Theta$ is not too small~\cite{SI}.
\bibitem{parameters} Unless otherwise stated parameters used are: 
$N=n=10$, $\alpha=100$, $\tau=10$ s, $v=100$ bp/s, $k_0=0.001$ s$^{-1}$, $D=2.25 \times 10^3$ bp$^2$/s,  while $\bar{J}$ is varied. Eq.~(\ref{model}) was solved by finite difference on a lattice of 
$L=1000$ sites, with $\Delta x=15$ bp and $\Delta t=0.01$ s (this mapping is relevant to the bacterial case). Technical details regarding the solution of Eq.~(\ref{model}) are given in \cite{SI}.
\end{thebibliography}
\end{document}